\documentclass[%
 reprint,
 amsmath,amssymb,
 aps,
]{revtex4-1}

\usepackage{graphicx}
\usepackage{dcolumn}
\usepackage{bm}
\usepackage{color}
\usepackage[normalem]{ulem}

\usepackage[normal]{subfigure}
\begin{document}


\title{  On the holographic phase transitions at finite topological charge}
\thanks{  On the holographic phase transitions at finite topological charge}%

\author{Tran Huu Phat}%
 \email{thphat@live.com}
\affiliation{%
 Vietnam Atomic Energy Commission, 59 Ly Thuong Kiet, Hanoi, Vietnam.}%


\author{Toan T. Nguyen}
\affiliation{Key Laboratory for Multiscale Simulation of Complex Systems, 
University of Science, Vietnam National  University, Hanoi \\
334 Nguyen Trai street, Thanh Xuan district, Hanoi 120000, Vietnam}
%


\date{\today}

\begin{abstract}
Exploring the significant impacts of topological charge on the holographic phase transitions and conductivity we start from an Einstein – Maxwell system coupled  with a charged scalar field in Anti – de Sitter spacetime. In our set up, the corresponding black hole (BH) is chosen to be the topological AdS one where the pressure is identified with the cosmological constant \cite{c12}. Our numerical computation shows that the process of condensation is favored at finite topological charge and, in particular, the pressure variation in the bulk generates a mechanism for changing the order of phase transitions in the boundary: the second order phase transitions occur at pressures higher than the critical pressure of the phase transition from small to large BHs while they become first order at lower pressures. This property is confirmed with the aid of holographic free energy. Finally, the frequency dependent conductivity exhibits a gap when the phase transition is second order and when the phase transition becomes first order this gap is either reduced or totally lost. \\
PACS numbers: 11.25.Tq , 04.70.Bw, 74.20.-z.  
\end{abstract}

\pacs{ 11.25.Tq , 04.70.Bw, 74.20.-z}
\maketitle


\section{Introduction}
It is known that the AdS/CFT duality \cite{c1} formulated by Maldacena and developed independently by Witten \cite{c2} and Gubser, Klebanov and Polyakov \cite{c3} in the form of the GKPW relation has opened up a new direction of connecting gravity 
with other branches of physics. 
At present the GKPW relation turns out to be a powerful formalism 
 for this purpose. 
In this respect, the study of holographic phase transitions has been developed 
strongly and gained great successes associated with superconductors and 
related topics \cite{c4, c5, c6, c7, c8, c9, c10, c11, c12, c13, c14, c15, c16, c17}. Following this trend we start from  the model of a Abelian Higgs field and a Maxwell field in the four – dimensional spacetime Einstein gravity. The bulk action reads

\begin{align}\label{q1}
S=&\frac{1}{16\pi {{G}_{N}}}\int{{{d}^{4}}x\sqrt{-g}}(R-\frac{6}{{{L}^{2}}}-\frac{1}{4}F_{\mu \nu }^{2} \\ \nonumber
& -{{\left| {{\partial }_{\mu }}-iq{{A}_{\mu }}\psi  \right|}^{2}}-m{{\left| \psi  \right|}^{2}}), \\ \nonumber 
\end{align}
where $G_{N}$ is the Newton constant. In the uncondensed phase, the solutions to Eq. (\ref{q1}) are the Reissner – Nordstrom black hole (BH)  
\begin{equation}\label{q2}
d{{s}^{2}}=-f\left( r \right)d{{t}^{2}}+\frac{d{{r}^{2}}}{f\left( r \right)}+{{r}^{2}}d\Omega _{2,k}^{2},
\end{equation}
where
\begin{equation}
f\left( r \right)=k-\frac{M}{r}+\frac{{{Q}^{2}}}{{{r}^{2}}}+\frac{{{r}^{2}}}{{{L}^{2}}},
\end{equation}
and
\begin{eqnarray*}
\phi (r) &=& Q\left( \frac{1}{r}-\frac{1}{{{r}_{0}}} \right) , \nonumber \\
\psi (r) &=& 0 , \nonumber
\end{eqnarray*}
here the horizon radius, $r_0$, is the largest solution of the equation:
\begin{equation}\label{q4}
f\left( {{r}_{0}} \right)=k-\frac{M}{{{r}_{0}}}+\frac{{{Q}^{2}}}{r_{0}^{2}}+\frac{r_{0}^{2}}{{{L}^{2}}}=0.
\end{equation}
In Eq. (\ref{q2}), $d\Omega _{2,k}^{2}$  is the metric of a two – sphere of radius $1/\sqrt{k}$  for  $k > 0$.  Note that the parameters $M$ and $Q$ are different from the mass and the charge of BH by corresponding factors.

The Hawking temperature $T$ and entropy of BH 
are respectively given by
\begin{subequations}
\begin{align}\label{q5}
& T=\frac{{f}'\left( {{r}_{0}} \right)}{4\pi }, \\ 
& S=\pi r_{0}^{2}\label{q5b}.
\end{align}
\end{subequations}
It is very interesting to mention that adopting the relation between pressure P and the cosmological constant  of BH 
\begin{equation}\label{q6}
P=-\frac{\Lambda }{8\pi }=\frac{3}{8\pi {{L}^{2}}},
\end{equation}
one discovered the total analogy between the small – large BH phase transition and the liquid – gas  phase transition of the van der Waals theory for $k > 0$ \cite{c11, c12} . In this set up $M$ becomes the enthalpy of the system and $k$ is interpreted as the measure of a new charge, the topological charge \cite{c13, c14}. Then the extended first law \cite{c15} of BH reads
\begin{equation*}
dM=TdS+\omega d\varepsilon +VdP+\phi dQ,
\end{equation*}
where $T$ and $S$ are given in Eq. (\ref{q5}),  $\varepsilon =4\pi k$ is the topological charge  and its conjugate potential $\omega = r_0/8\pi$, the pressure $P$ conjugates to the volume  $V=\frac{4}{3}\pi r_{0}^{3}$, the charge $Q$ conjugates 
to the potential $\phi = Q/r_0$.

In term of $\varepsilon$ from Eqs. (\ref{q4}), (\ref{q5b}) and (\ref{q6}), it is easily derived the isobaric specific heat 
\begin{equation*}
{{C}_{PQ\varepsilon }}=T{{\left( \frac{\partial S}{\partial T} \right)}_{PQ\varepsilon }}=\frac{2\pi r_{0}^{2}\left( 32{{\pi }^{2}}\Pr _{0}^{4}+\varepsilon r_{0}^{2}-4\pi {{Q}^{2}} \right)}{32{{\pi }^{2}}\Pr _{0}^{4}-\varepsilon r_{0}^{2}+12\pi {{Q}^{2}}},
\end{equation*}
which yields the critical pressure for the phase transition from small to large BHs
\begin{equation}\label{q8}
{{P}_{c}}=\frac{{{\varepsilon }^{2}}}{1536{{\pi }^{3}}{{Q}^{2}}} ~,~ {{T}_{c}}=\frac{\sqrt{6}{{\varepsilon }^{3/2}}}{96{{\pi }^{3/2}}Q} ~,~
{{r}_{c}}=\frac{2\sqrt{6\pi }Q}{\sqrt{\varepsilon }}~.
\end{equation}

It is worth to emphasize that in Ref. \cite{c16} the authors proved that the action Eq. (\ref{q1}) together with Eq. (\ref{q2}) yields the spontaneous breaking of gauge invariance at finite $\varepsilon$. Inspired by this result, we will investigate systematically in this paper the holographic phase transitions at finite topological charges. Our main aim is to look for the new effect which could occur when the pressure $P$ varies from $P>{{P}_{c}}$  to $P\leq{{P}_{c}}$ in the bulk. For simplicity we set $Q =1$ from now on. 

The present paper is organized as follows. Section \ref{2} deals with the holographic phase transitions at finite topological charges and the free energy, respectively. Section \ref{3} is devoted to the calculations of frequency dependent conductivity in different phase transitions. The conclusion and outlook are given in Section \ref{4}.

\section{Holographic phase transitions }\label{2}

\subsection{Basic set up}

At first let us set up the frame work for the whole study  of holographic phase transitions. We begin with the following ansatz 
\begin{align}\label{q21}
		{{A}_{\mu }}=\left( \phi \left( r \right),0,0,0 \right),\psi =\psi \left( r \right),
\end{align}

and at the same time we choose 
\begin{equation}\label{q22}
{{m}^{2}}=-\frac{2}{{{L}^{2}}},
\end{equation}
which is above the Breitenlohner – Freedman bound \cite{c18}.

Inserting Eq. (\ref{q21}) into Eq. (\ref{q1}) we derive the equations of motion 
for matter fields 

\begin{align}
& {\phi }''+\frac{2}{r}+{\phi }'-\frac{2{{\psi }^{2}}}{f}\phi =0,\label{q23} \\ 
& {\psi }''+\left( \frac{2}{r}+\frac{{{f}'}}{f} \right){\psi }'+\left( \frac{{{\phi }^{2}}}{{{f}^{2}}}+\frac{2}{{{L}^{2}}f} \right)\psi =0,\label{q24}
\end{align}
where the prime denotes derivative with respect to $r$. 

For the fields to be regular at horizon we impose the condition
\begin{equation}\label{q25}
\phi \left( {{r}_{0}} \right)=0 .
\end{equation}
Inserting Eq. (\ref{q25}) into Eq. (\ref{q24}), and
expand near $r\rightarrow r_0$,
we arrive at the condition at horizon for scalar field
\begin{equation}\label{q26}
\frac{{\psi }'\left( {{r}_{0}} \right)}{\psi \left( {{r}_{0}} \right)}=-\frac{1}{2\pi {{L}^{2}}T}.
\end{equation}
At the AdS boundary, the large $r$  behaviors of $\phi$ and $\psi $ take the form
\begin{align}
& \psi \left( r \right)=\frac{{{\psi }_{1}}}{r}+\frac{{{\psi }_{2}}}{{{r}^{2}}}+...\label{q27} \\ 
& \phi \left( r \right)=\mu -\frac{\rho }{r}+...,\label{q28}  
\end{align}
where $\mu $ and $\rho $ are chemical potential and  the corresponding density associated with the expectation value of charge density, $\rho =\left\langle {{J}^{0}} \right\rangle $, with a source term in the boundary action of the form
\begin{equation*}
{{S}_{bdy}}\to {{S}_{bdy}}+\mu \int{{{d}^{4}}}x{{J}^{0}}\left( x \right).
\end{equation*} 
Due to the holographic duality there are two possibilities for identifying the sources and condensates of the dual field theory

\begin{itemize}
\item ${{\psi }_{1}}$  is the source which vanishes at infinity
\begin{equation}\label{q29}
{{\psi }_{1}}=0,
\end{equation}
and ${{\psi }_{2}}$ is condensate ${{\psi }_{2}}\sim \left\langle {{O}_{2}} \right\rangle $.\\

\item	${{\psi }_{2}}$  is the source which vanishes at infinity
\begin{equation}\label{q210}
{{\psi }_{2}}=0,                                                             
\end{equation}
and ${{\psi }_{1}}$  is condensate  ${{\psi }_{1}}\sim \left\langle {{O}_{1}} \right\rangle$.
\end{itemize}

\subsection{Free enrergy }
In order to analyse the order of phase transitions  we have to calculate the holographic free energy. This quantity is holographically evaluated by calculating the corresponding on-shell value of the Abelian – Higgs sector of the Euclidean action. Plugging the ansatz, Eq. (\ref{q21}), into the action, Eq. (\ref{q1}), we arrive at  
\begin{equation}\label{q211}
{{S}_{M}}=\int{{{d}^{4}}x\left[ \frac{{{{{\phi }'}}^{2}}\left( z \right)}{2}-f\left( z \right){{{{\psi }'}}^{2}}\left( z \right)+\frac{{{\phi }^{2}}\left( z \right){{\psi }^{2}}\left( z \right)}{{{z}^{4}}f\left( z \right)}+\frac{2{{\psi }^{2}}\left( z \right)}{{{z}^{4}}} \right]}.
\end{equation}
Here we have made a change of variable, $z=1/r$, to better formulate
the system of equations of motion for numerical evaluations and to simplify various expressions.

Applying the boundary condition, Eq. (\ref{q25}), and the Eqs. (\ref{q23}, \ref{q24}),
we obtain the on – shell value of the Euclidean action
\begin{align*}
& {{S}_{OS}}=\int{\frac{1}{k}\left( {\left. \frac{{\phi }'(z)\psi (z)}{z^2} \right|}_{z=0}
- {{\left. \frac{f(z){\psi }'(z)\psi(z)}{z} \right|}_{z=0}} \right)}{{d}^{3}}x \\ 
& -\int{\frac{1}{k}\left( \int\limits_{0}^{{{z}_{0}}}{dz\frac{{{\phi }^{2}}{(z)}
\psi ( z)}{{{z}^{4}}f\left( z \right)}} \right)}{{d}^{3}}x.
\end{align*}

Substituting the asymptotic behaviors of  $\phi $ and $\psi $ into the above action we get
\begin{align*}
& {{S}_{OS}}=\int{\frac{1}{k}\left( \frac{\mu \rho }{2}+\frac{3{{\psi }_{1}}{{\psi }_{2}}}{{{L}^{2}}}+{{\left( \frac{\psi _{1}^{2}}{z{{L}^{2}}} \right)}_{z=0}} \right){{d}^{3}}}x \\ 
& -\int{\frac{1}{k}}\left( \int\limits_{0}^{{{z}_{0}}}{dz\frac{{{\phi }^{2}}\left( z \right){{\psi }^{2}}\left( z \right)}{{{z}^{4}}f\left( z \right)}} \right){{d}^{3}}x. \\ 
\end{align*}

The divergence term in the foregoing expression will be removed by adding the counter term \cite{c8}
\begin{equation*}
{{S}_{c}}=-\frac{1}{k{{L}^{2}}}{{\int{{{d}^{3}}x\left( \sqrt{-h}{{\psi }^{2}}(z) \right)}}_{z=0}},
\end{equation*}
where $ h $ is the determinant of the induced metric on the AdS boundary. With the aid of the asymptotic behavior of $\psi$ it is  easily found that
\begin{equation*}
{{S}_{c}}=-\frac{1}{k{{L}^{2}}}\int{{{d}^{3}}x\left[ 2{{\psi }_{1}}{{\psi }_{2}}+{{\left( \frac{\psi _{1}^{2}}{z} \right)}_{z=0}} \right]}.
\end{equation*}
The renormalised free energy of the boundary field theory  is obtained 
\begin{align}\label{q212}
& {{\Omega }_{1}}=-T\left( {{S}_{OS}}+{{S}_{c}} \right) \\ \nonumber
& ={{V}_{2}}\left[ -\frac{\mu \rho }{2}-\frac{{{O}_{1}}{{O}_{2}}}{{{L}^{2}}}+\int\limits_{0}^{{{z}_{0}}}{dz\frac{{{\psi }^{2}}\left( z \right){{\phi }^{2}}\left( z \right)}{{{z}^{4}}f\left( z \right)}} \right], 
\end{align}
which corresponds to the ${{O}_{1}}$ quantization ,  ${{V}_{2}}$ is the volume of two – sphere with radius  $1/\sqrt{k}$.

Analogously, the renormalised free energy in ${{O}_{2}}$ quantization reads
\begin{equation}\label{q213}
{{\Omega }_{2}}={{V}_{2}}\left[ -\frac{\mu \rho }{2}+\frac{{{O}_{1}}{{O}_{2}}}{{{L}^{2}}}+\int\limits_{0}^{{{z}_{0}}}{dz\frac{{{\psi }^{2}}\left( z \right){{\phi }^{2}}\left( z \right)}{{{z}^{4}}f\left( z \right)}} \right],
\end{equation}
which is the Legendre transform of ${{\Omega }_{1}}$. From Eq. (\ref{q212}) and Eq. (\ref{q213}),  it is clear that
\begin{equation*}
\frac{1}{{{V}_{2}}}\frac{\partial {{\Omega }_{1}}}{\partial {{O}_{1}}}=-{{O}_{2}},\quad 
\frac{1}{{{V}_{2}}}\frac{\partial {{\Omega }_{2}}}{\partial {{O}_{2}}}={{O}_{1}},
\end{equation*} 
implying that the local extrema of ${{\Omega }_{1}}({{\Omega }_{2}})$ locates at vanishing ${{O}_{2}}\left( {{O}_{1}} \right)$. This is exactly what we assumed that only one of  
${{\psi }_{1}}$ and ${{\psi }_{2}}$ is non-vanishing for physical solutions.

The free energy corresponding to non-condensed state reads
\begin{equation}\label{q214}
\frac{{{\Omega }_{0}}}{{{V}_{2}}}=\frac{{{\mu }^{2}}{{z}_{0}}}{2}.
\end{equation}
From Eq. (\ref{q214}), we get the free energy difference
\begin{equation}\label{q215}
\frac{\Delta {{\Omega }_{i}}}{{{V}_{2}}}=\frac{{{\Omega }_{i}}}{{{V}_{2}}}-\frac{\mu^2 {{z}_{0}}}{2}
\quad\quad\left( \text{ }i\text{ }=\text{ }1\text{ },\text{ }2\text{ } \right).
\end{equation}
This is our expected result.

\subsection{Numerical results}

In this subsection we focus on the impacts of the topological charge in the holographic phase transitions. To this end, let us proceed to the numerical calculation method 
which will be implemented in the following cases:

\subsubsection{$\varepsilon =4\pi, L =  1 $}

\begin{figure}
\centering
	\subfigure[]{\includegraphics[width=0.45\textwidth]{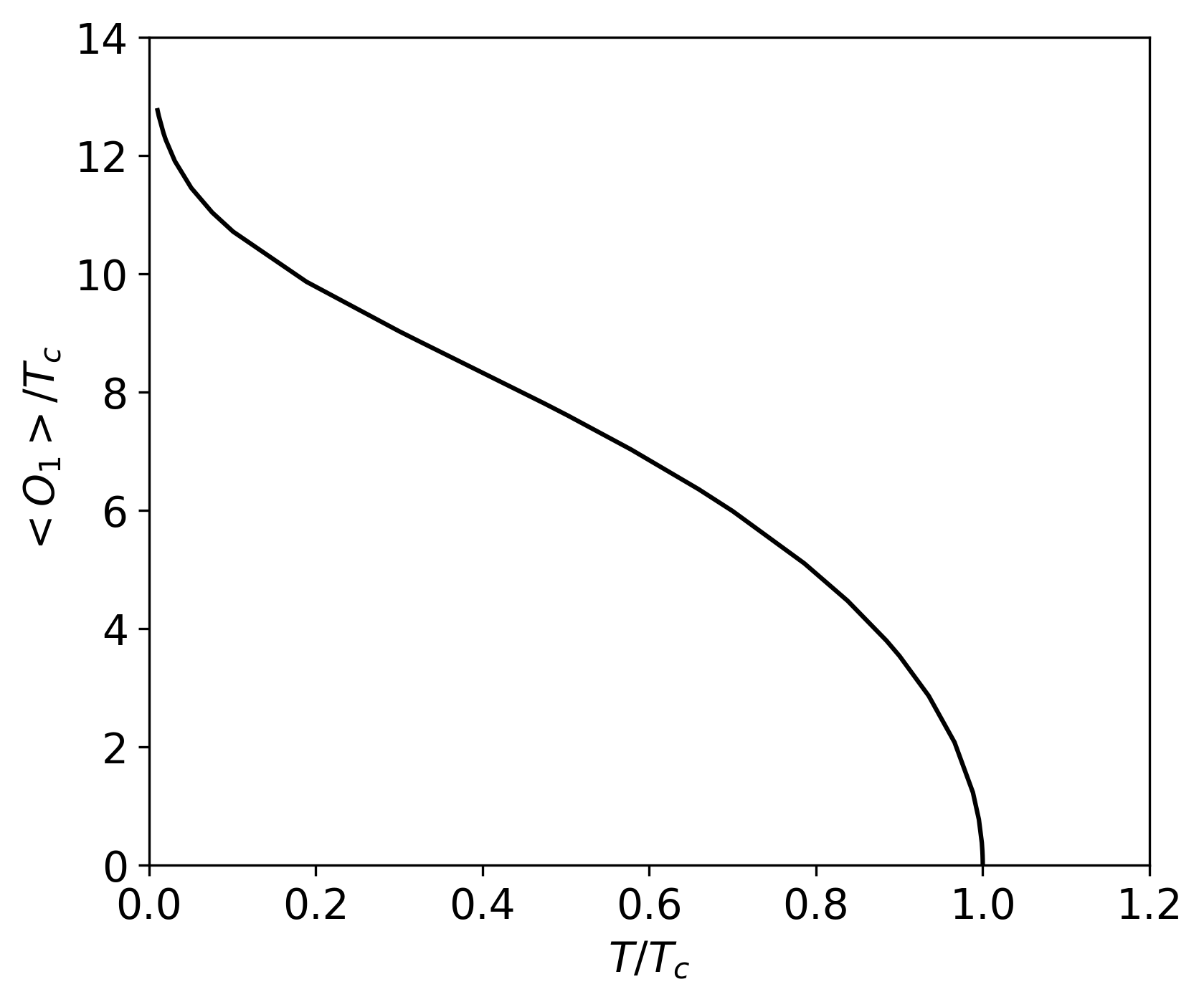}
		\label{subfig:1a}}
	\subfigure[]{\includegraphics[width=0.45\textwidth]{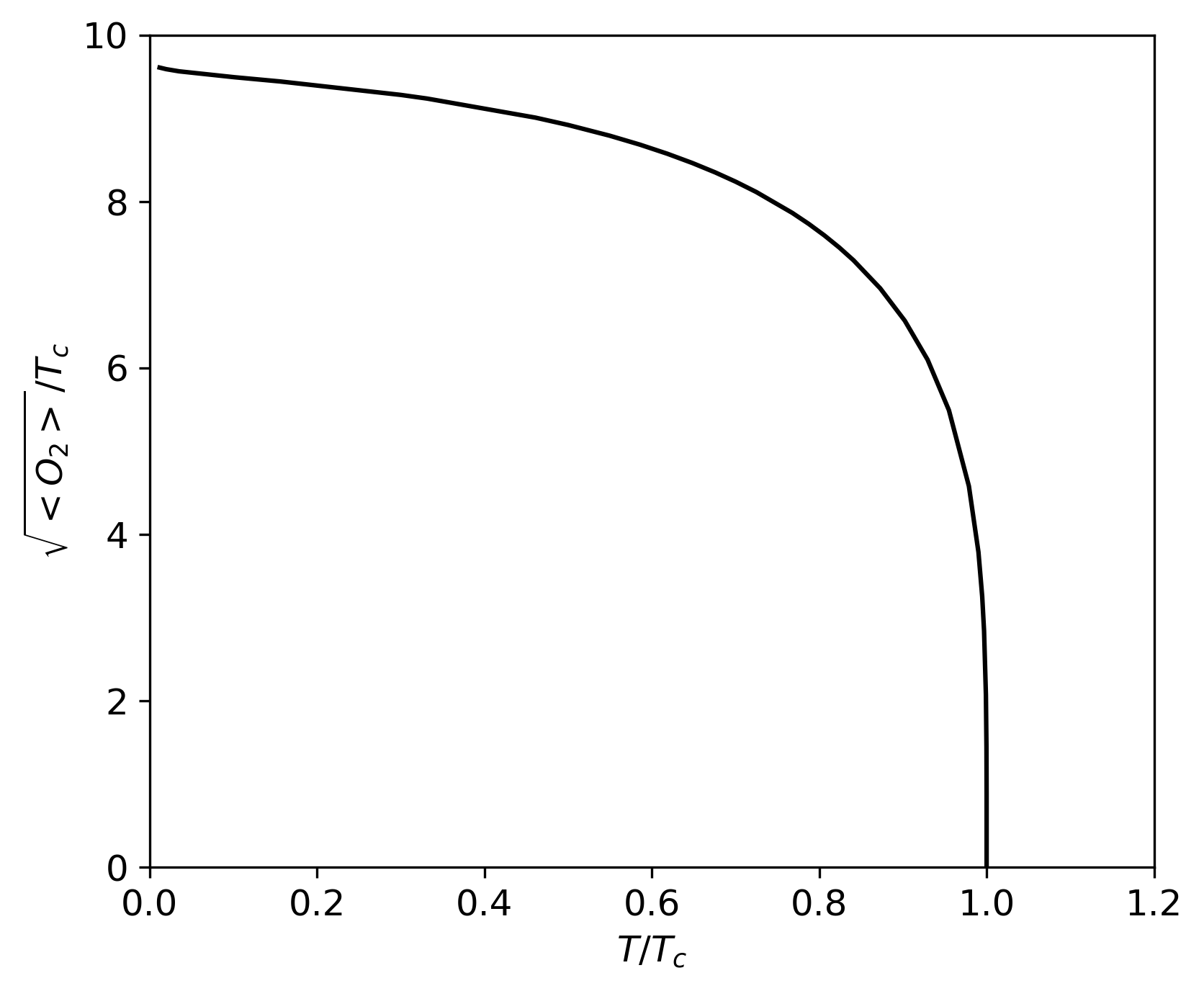}
		\label{subfig:1b}}
\caption{The condensation of  $\left\rangle O_1\right\rangle $ (a) and 
$\left\rangle O_2\right\rangle$ (b).}
\label{fig:L1}
\end{figure}

The scalar condensation is plotted in Figs. \ref{fig:L1},  where the onset of second order phase transitions 
occur at  ${{T}_{c}}=0.238{{\rho}^{1/2}},0.046{{\rho }^{1/2}}$, respectively.
After fitting the curves in Figs. \ref{fig:L1} 
near $T\rightarrow T_c$, we obtain approximately the expressions
\begin{equation*}
\frac{\left\langle {{O}_{1}} \right\rangle}{T_c} 
    = \frac{\sqrt{2}{{\psi }_{1}}}{T_c} \approx 11.318~ {
    \left( 1-\frac{T}{T_c} \right)}^{1/2},
\end{equation*}
as 
$T\rightarrow T_c=0.113\mu$
and
\begin{equation*}
\frac{\sqrt{\left \langle O_2 \right\rangle}}{T_c} 
 = \frac{\sqrt{\sqrt{2}{\psi }_{2}}}{T_c}
 \approx 32.966~ {{\left( 1-\frac{T}{T_c} \right)}^{1/2}} .
\end{equation*}
as 
$T \rightarrow T_c=0.050\mu$.

Figs. \ref{fig:L1} and the foregoing  expressions characterise one of the typical 
properties of superconductors in mean-field approximation. 
It is worth to note that the phase transition takes place at $ L= 1 $, corresponding to pressure $ {{P}_{L=1}}={{\left( \frac{3}{8\pi {{L}^{2}}} \right)}_{L=1}}$ which is bigger than the critical pressure ${{P}_{c}}={{\left( \frac{3}{8\pi {{L}^{2}}} \right)}_{L=6}}$ given in 
Eq. (\ref{q8}).

\begin{figure}
\centering
	\subfigure[]{\includegraphics[width=0.45\textwidth]{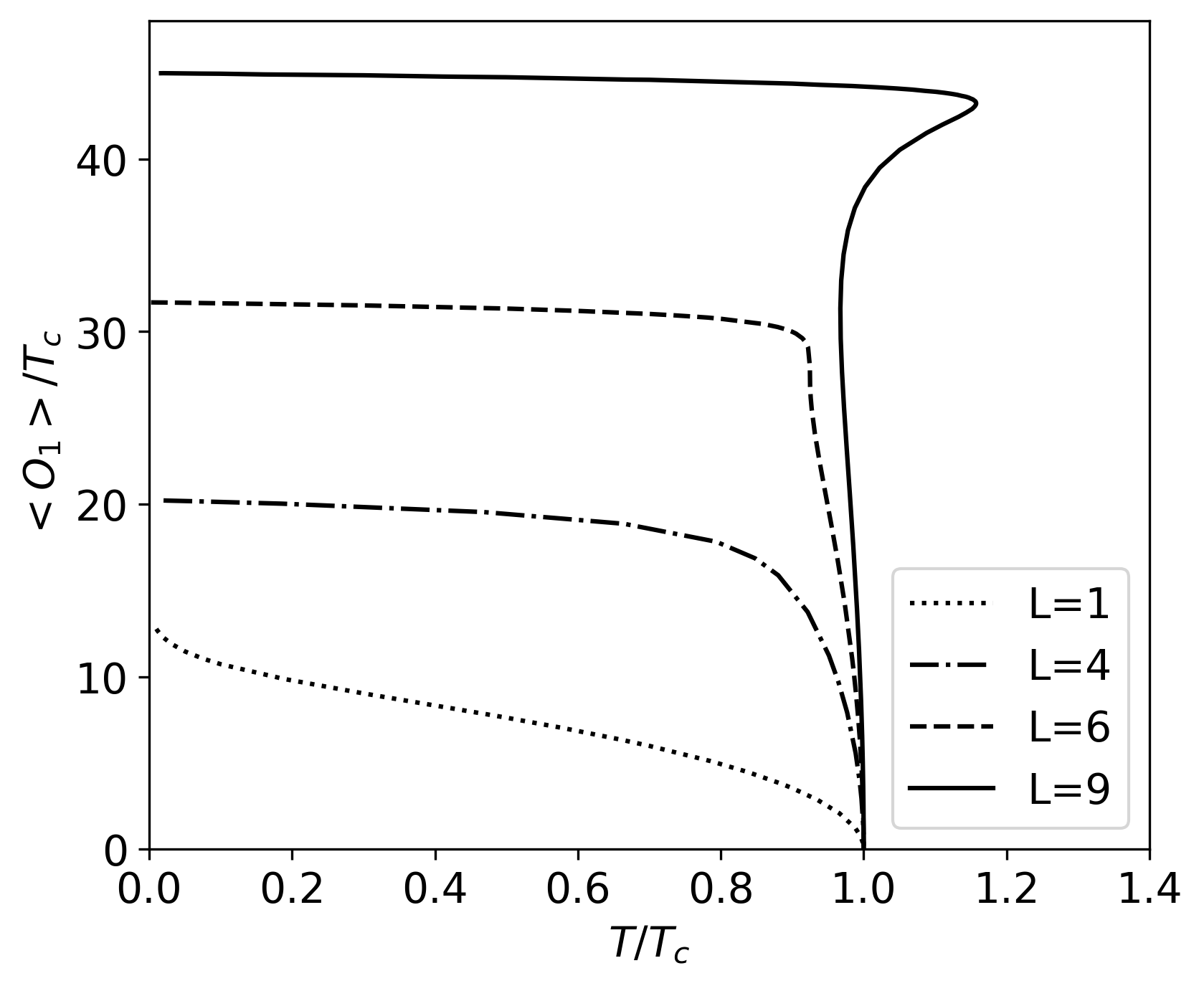}
		\label{subfig:2a}}
	\subfigure[]{\includegraphics[width=0.45\textwidth]{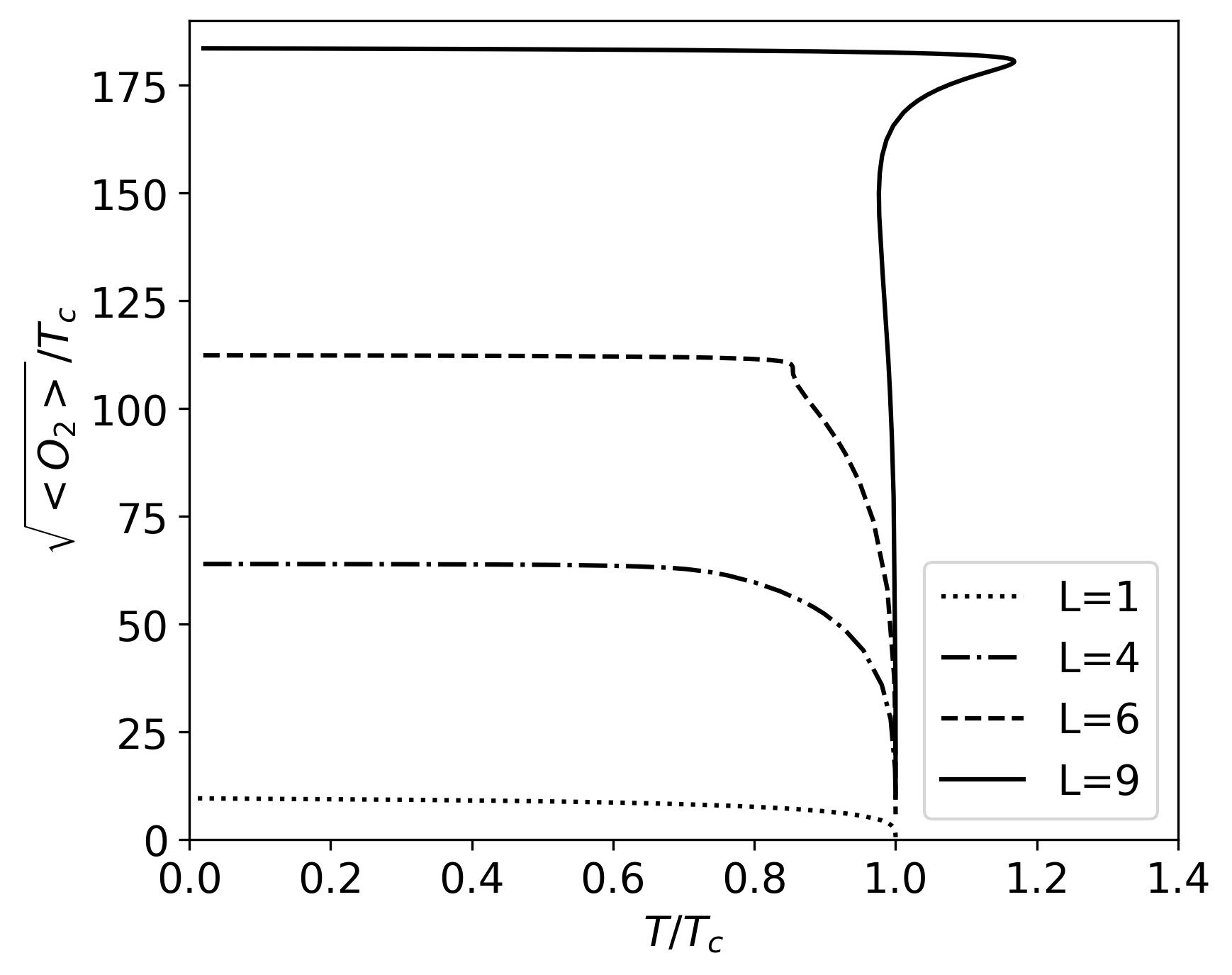}
		\label{subfig:2b}}
\caption{For $ L < 6 $ the phase transitions are second order while for $ L\geq 6 $ the first order phase transitions begin to exhibit for  $ O_1 $ (up) and $ O_2 $ (down).}
\label{fig:L1to9k1}
\end{figure}

\begin{figure}
\centering
	\subfigure[]{\includegraphics[width=0.45\textwidth]{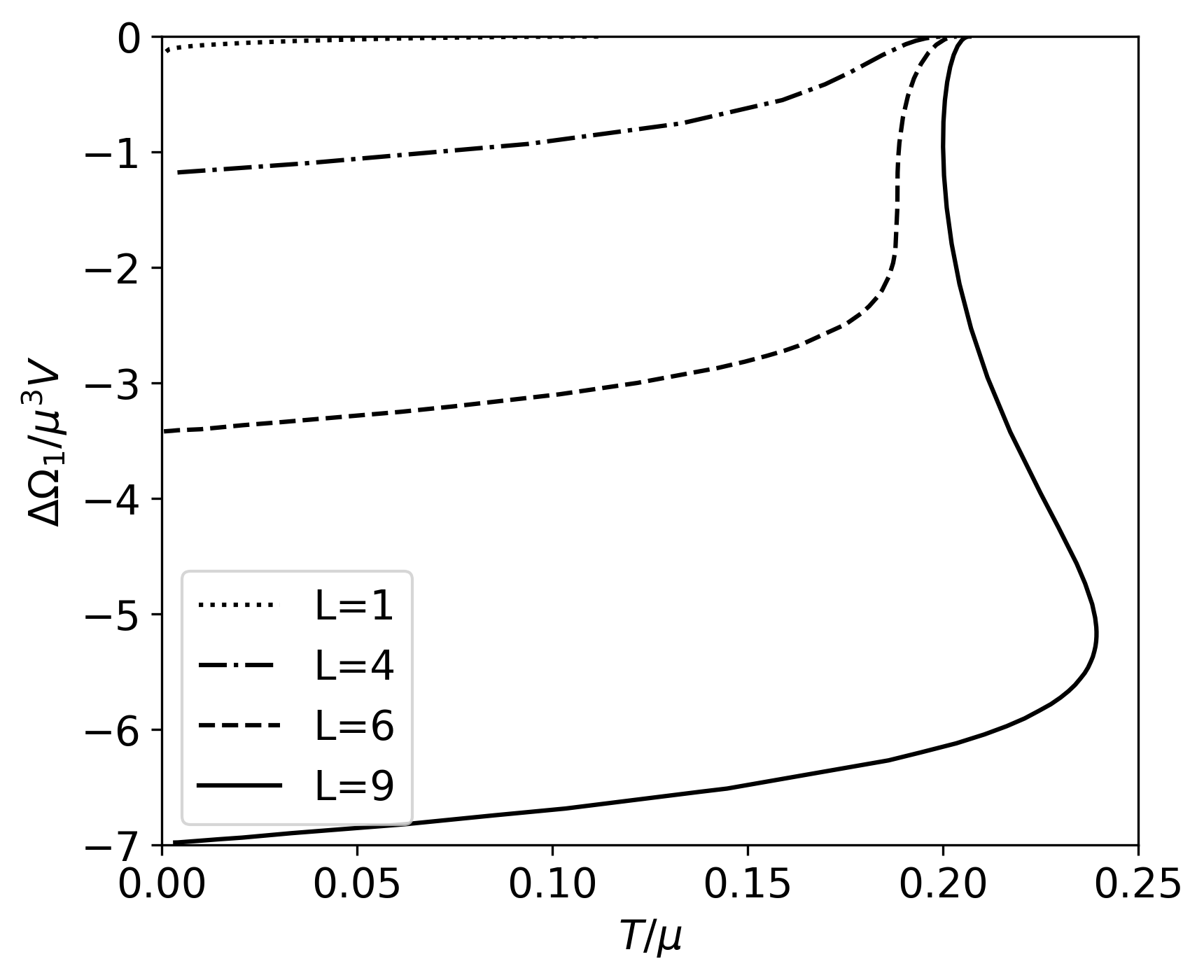}
		\label{subfig:3a}}
	\subfigure[]{\includegraphics[width=0.45\textwidth]{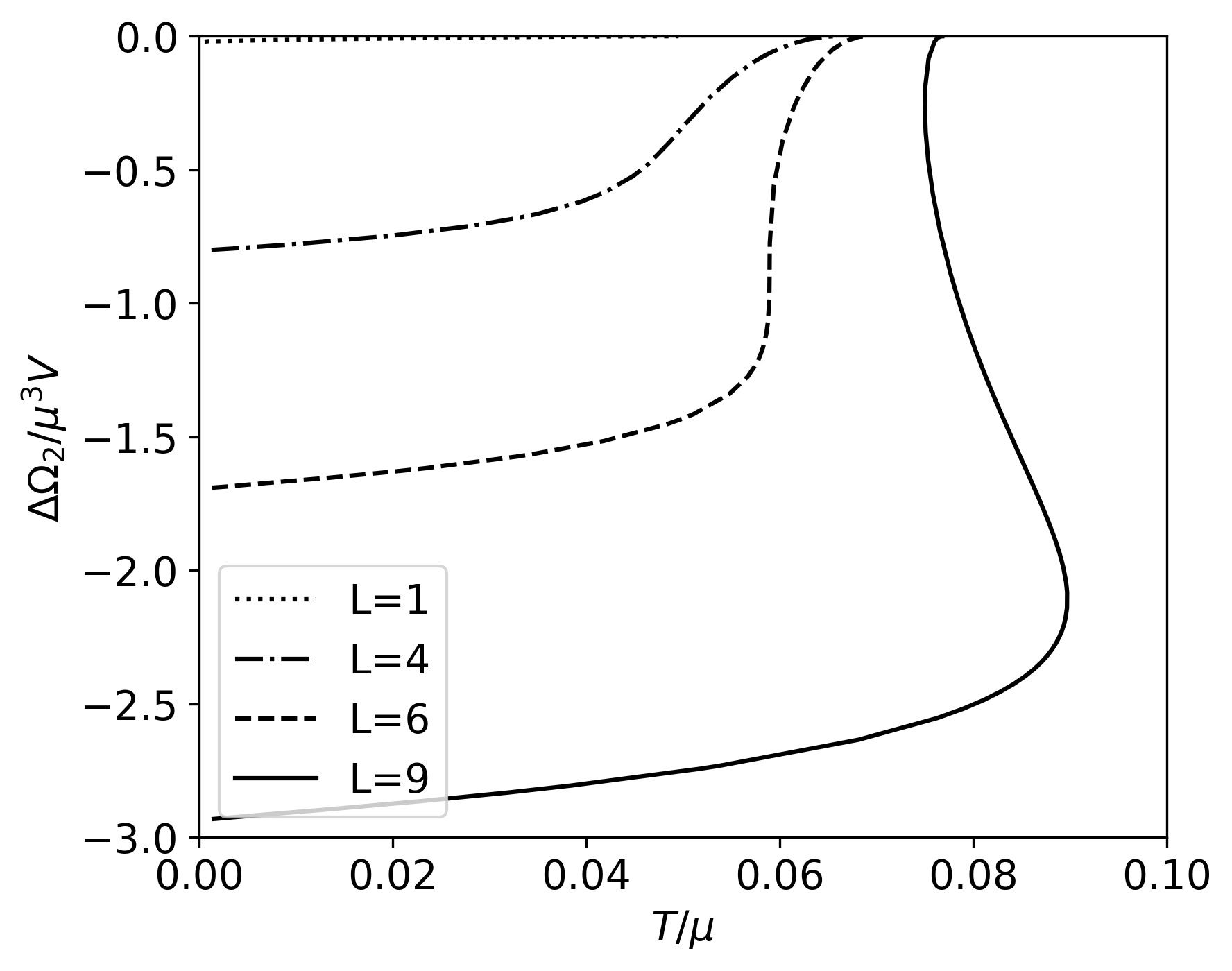}
		\label{subfig:3b}}
\caption{The free energy difference is not analytical at corresponding critical temperatures. 
Fig. 3a (Fig.3b) corresponds to $ O_1 $  ($ O_2 $).}
\label{fig:FEL1to9k1}
\end{figure}

Now let the pressure decrease to lower values which  corresponds to bigger values of $L$. 
In Figs. \ref{subfig:2a},
\ref{subfig:2b} are shown the phase diagrams for $L = $1, 4, 6, and 9. It is clear that at  critical pressure ${{P}_{c}}={{\left( \frac{3}{8\pi {{L}^{2}}} \right)}_{L=6}}$, the first order phase transition begins to appear.
In order to confirm exactly the exhibition of first order phase transitions let us calculate numerically the free energy difference, Eq. (\ref{q215}), between the condensed and uncondensed phases for $ O_1 $ and $ O_2 $, respectively. They are plotted in 
Fig. \ref{subfig:3a}, \ref{subfig:3b} which indicate that at $ L = 6$, 9 
the free energy difference is not analytical at corresponding critical temperatures.

\subsubsection{$\varepsilon =16\pi $  and $ L = 0.5,~ 1.5,~ 2$}

\begin{figure}
\centering
	\subfigure[]{\includegraphics[width=0.45\textwidth]{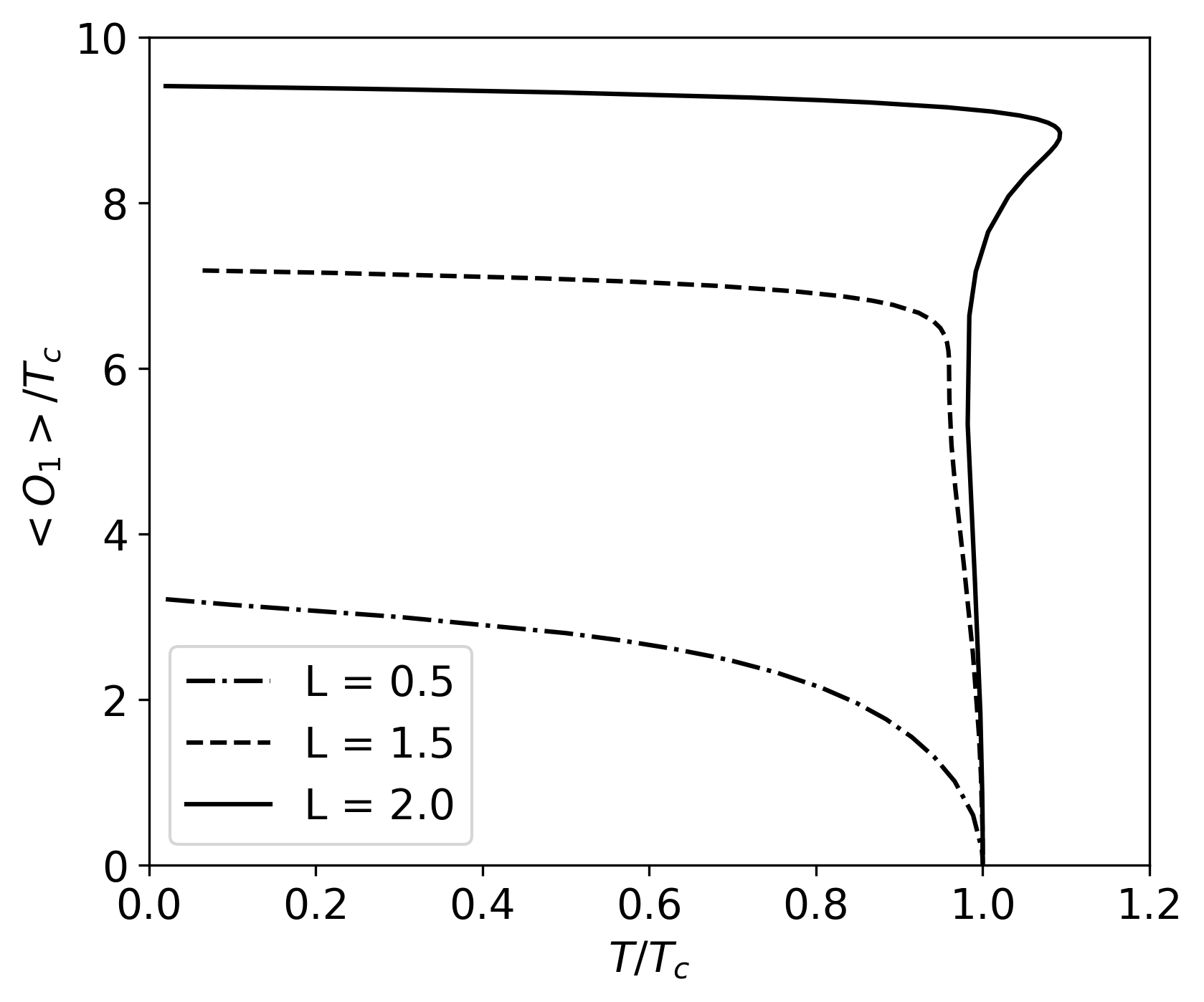}
		\label{subfig:4a}}
	\subfigure[]{\includegraphics[width=0.45\textwidth]{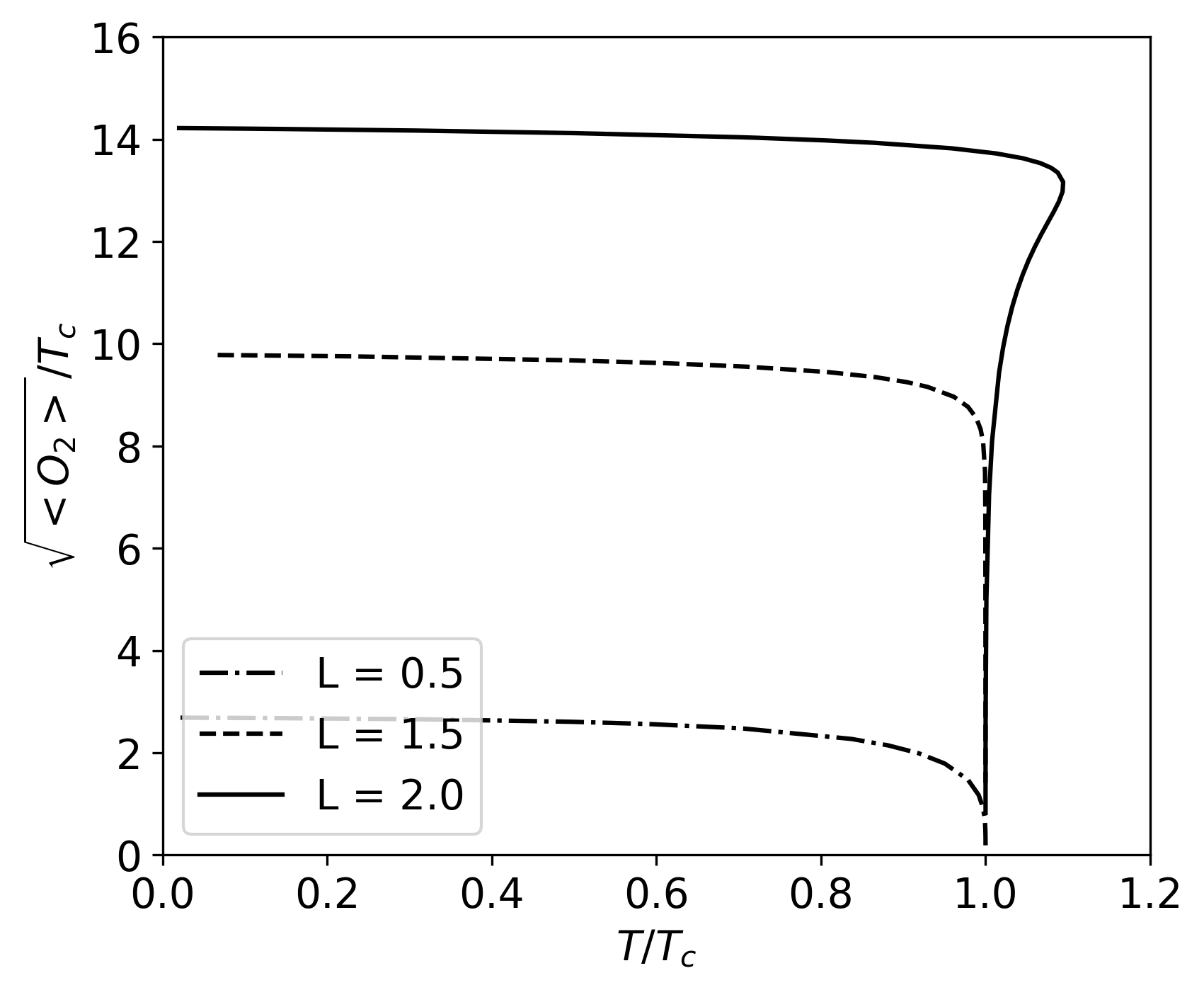}
		\label{subfig:4b}}
\caption{For $ L < 1.5$  phase transition is second order. It turns out to be first order  at $L = 1.5$. See $ O_1 $ (up) and $ O_2 $ (down).}
\label{fig:L1to2k4}
\end{figure}

\begin{figure}
\centering
	\subfigure[]{\includegraphics[width=0.45\textwidth]{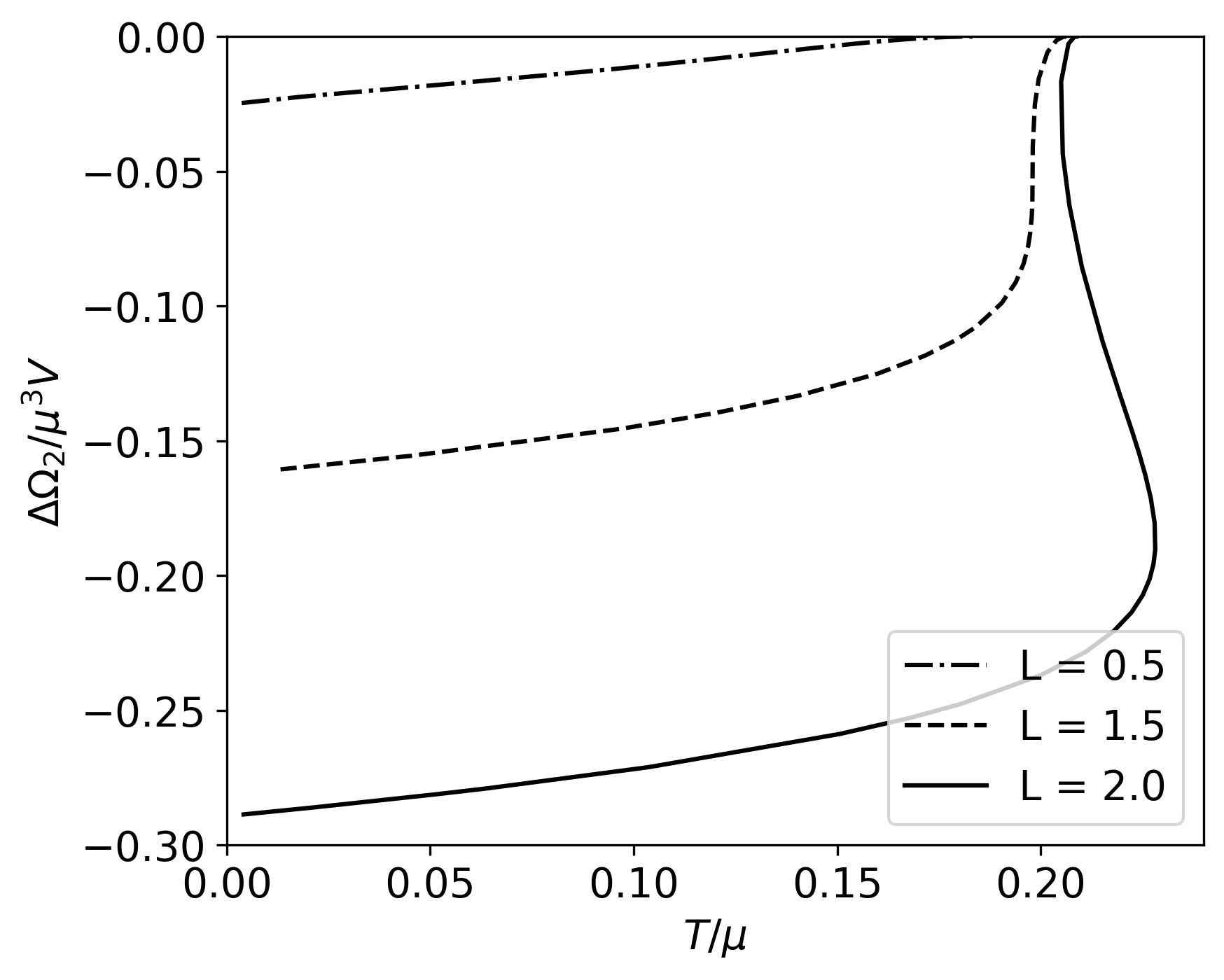}
		\label{subfig:5a}}
	\subfigure[]{\includegraphics[width=0.45\textwidth]{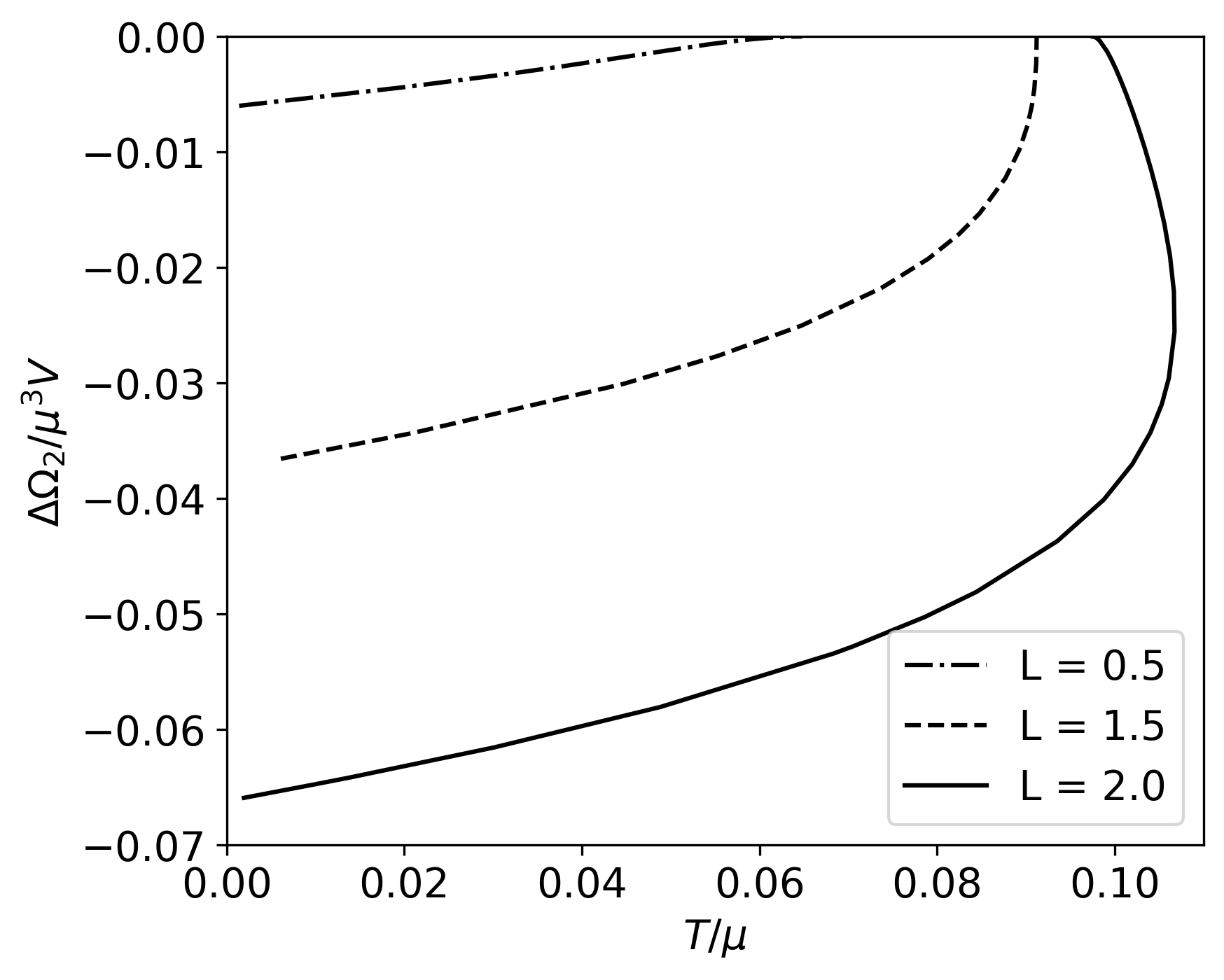}
		\label{subfig:5b}}
\caption{The free energy difference  calculated for $ O_1 $ quantization (up) and $ O_2 $ quantization (down) it is not analytical at corresponding critical temperatures.}
\label{fig:FEL1to2k4}
\end{figure}

Next we consider the case when the topological charge takes bigger value.
The phase diagrams plotted In Figs. \ref{fig:L1to2k4} tell that 
at $L\ge 1.5$ the first order phase transitions emerge.
Figs. \ref{fig:L1to2k4} demonstrate again that the first order phase transitions always emerge at critical pressure ${{P}_{c}}={{\left( \frac{3}{8\pi {{L}^{2}}} \right)}_{L=1,5}}$ and moreover when the topological charge gets bigger values the first order phase transitions manifest at higher pressures. The existence of the first order phase transitions in Figs. \ref{fig:L1to2k4} are confirmed again in Figs. \ref{fig:FEL1to2k4} the free energy is not analytical at corresponding temperatures.

\section{Conductivity}\label{3}
Let us finally proceed to the conductivity of superconductor in the dual CFT as a function of frequency. For this purpose, we must solve the equation for the fluctuations of the vector potential $ {{A}_{x}}\left( r,t \right) $ in the bulk. Let this potential take the form 
\begin{equation*}
{{A}_{x}}\left( r,t \right)={{A}_{x}}\left( r \right)\exp \left( -i\omega t \right),
\end{equation*}
from which we get the equation of ${{A}_{x}}\left( r \right)$ in our set up
\begin{equation}\label{q31}
{{{A}''}_{x}}\left( r \right)+\frac{{{f}'}}{f}{{{A}'}_{x}}\left( r \right)+\left( \frac{{{\omega }^{2}}}{{{f}^{2}}}-\frac{2{{\psi }^{2}}}{f} \right){{A}_{x}}\left( r \right)=0,
\end{equation}
which is rewritten in new variable $ z = 1/r $ as
\begin{equation}\label{q32}
{{{A}''}_{x}}\left( z \right)+\left( \frac{2}{z}+\frac{{{f}'}}{f} \right){{{A}'}_{x}}\left( z \right)+\left( \frac{{{\omega }^{2}}}{{{z}^{4}}{{f}^{2}}}-\frac{2{{\psi }^{2}}}{{{z}^{4}}f} \right){{A}_{x}}\left( z \right)=0,
\end{equation}
where $\psi $ was solved in Section \ref{2}. In order to determine the solution to Eq.
(\ref{q31}, \ref{q32}) we need two boundary conditions. 
The first one is given by the on going condition at horizon 
\begin{equation}\label{q33}
{{A}_{x}}\left( r \right)\sim {{\left( r-{{r}_{0}} \right)}^{-i\omega /4\pi T}}+...\,
\end{equation}  
 and the second boundary condition is set at large  $ r $. To do this, $A_x$ is expanded  in term of $z$ near $z\rightarrow 0$. This gives
\begin{equation}\label{q34}
{{A}_{x}}\left( z \right)={{a}_{x}}+z{{b}_{x}}+...
\end{equation}
Therefore the second boundary condition is 
\begin{equation}\label{q35}
{{A}_{x}}\left( 0 \right)={{a}_{x}}.
\end{equation}
The AdS/CFT duality dictionary tells that ${{b}_{x}}$ determines the boundary current  ${{j}_{x}}={{b}_{x}}$. Then the Ohm law gives us the conductivity
\begin{equation}\label{q36}
\sigma \left( \omega  \right)=\frac{{{j}_{x}}}{{{E}_{x}}}=-\frac{i{{b}_{x}}}{\omega {{a}_{x}}}.
\end{equation}
Eq. (\ref{q36}) requires us to solve the differential equation (\ref{q31}, \ref{q32})  based on two boundary conditions (\ref{q33}) and (\ref{q34}).The numerical computation provides the graphs of the frequency dependent conductivity 
corresponding respectively to the phase diagrams given in subsection $C$. 
They are presented in accordance with different items in subsection $ C $:

\subsubsection{$\varepsilon = 4\pi, Q=1, L=1$, and $9$}

We plot the real part of frequency dependent conductivity Re$\sigma \left( \omega  \right)$ derived from the numerical calculations of Eqs. (\ref{q31}, \ref{q32}) at $ Q = k = L = 1 $ in Figs. 6a, 6b which exhibit a gap determined by condensate and, furthermore Re$\sigma \left( \omega  \right)$ contains a delta function which is recognized from the imaginary part Im$\sigma \left( \omega  \right)$ having a pole at $\omega =0$ (it is not shows up here). Therefore the real part and imaginary part of $\omega =0$ are related by the Kramers – Kronig relations. At $ L = 9 $ we have correspondingly Figs.7a, 7b which tell that the foregoing gap disappears. This means that the superconductivity is totally lost.
\begin{figure}
\centering
	\subfigure[]{\includegraphics[width=0.45\textwidth]{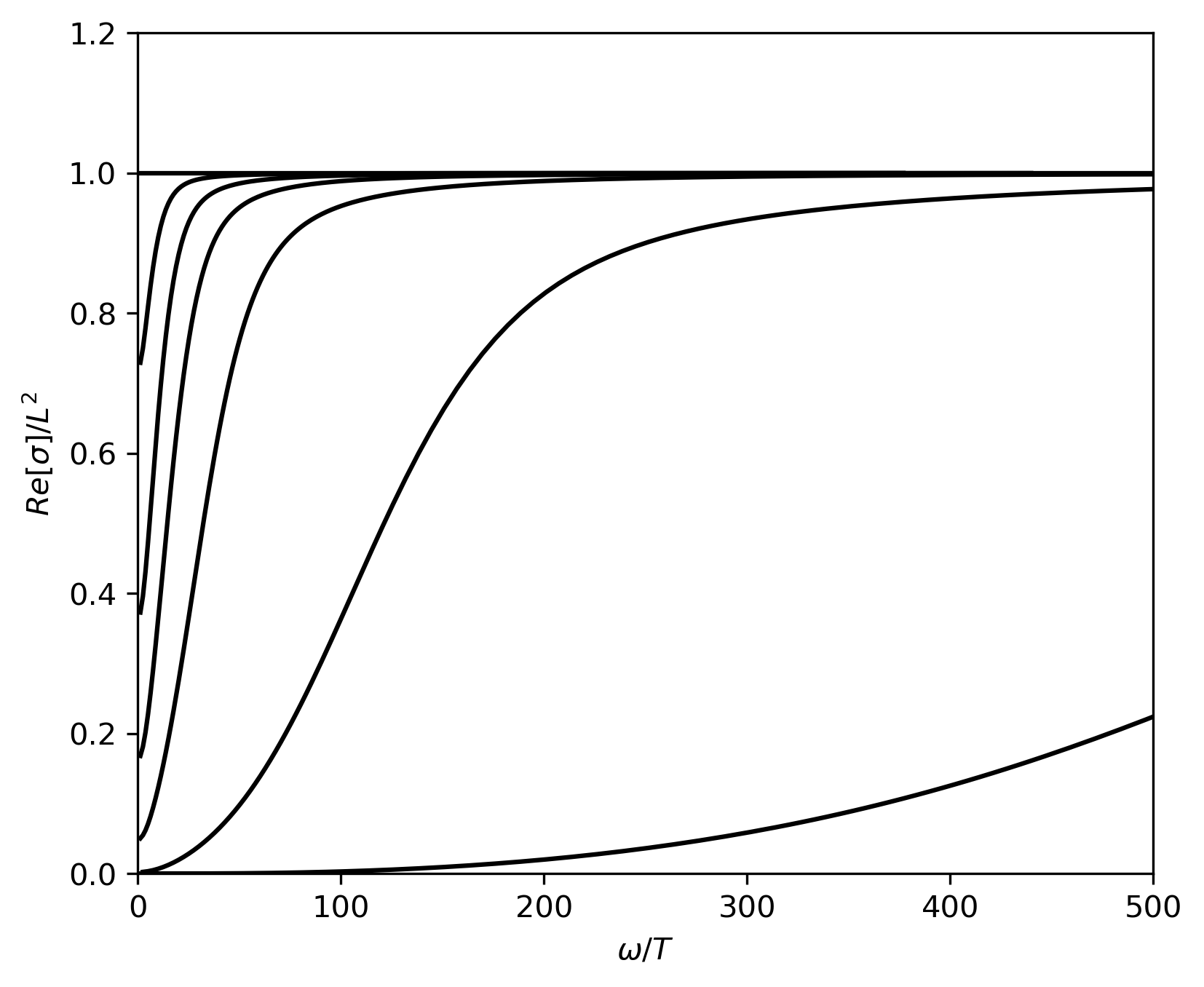}
		\label{subfig:6a}}
	\subfigure[]{\includegraphics[width=0.45\textwidth]{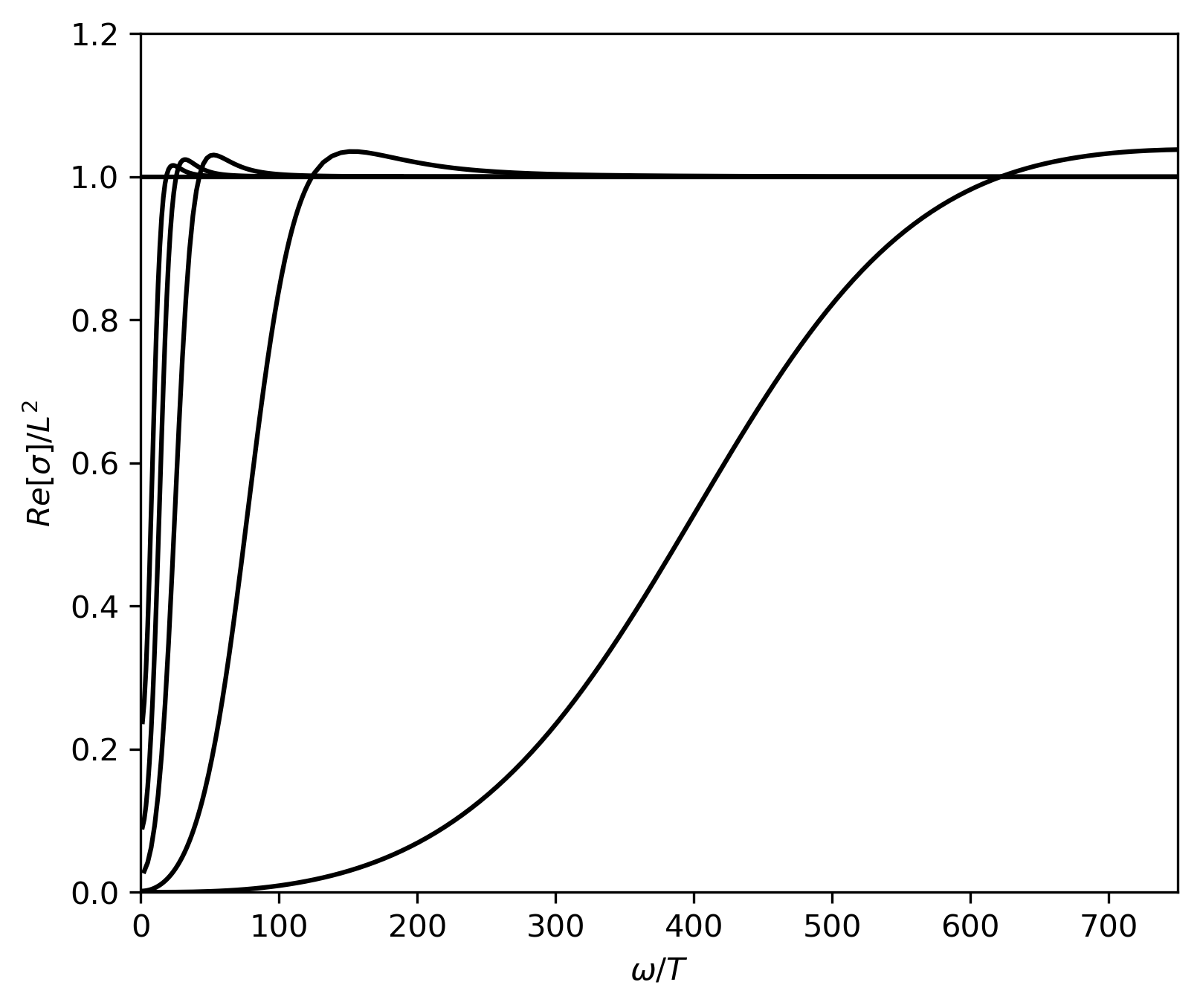}
		\label{subfig:6b}}
\caption{The evolution of Re $\sigma \left( \omega  \right)$  versus   $\omega /T$ at $\varepsilon =4\pi, Q = L = 1$  for $ O_1 $ (a) and $ O_2 $ (b). 
Each graph, from left to right, is plotted for $T/T_c$ = 0.9, 0.7, 0.5, 0.3, 0.1 and 0.02 respectively.}
\label{fig:sigmak1L1}
\end{figure}
\begin{figure}
\centering
	\subfigure[]{\includegraphics[width=0.45\textwidth]{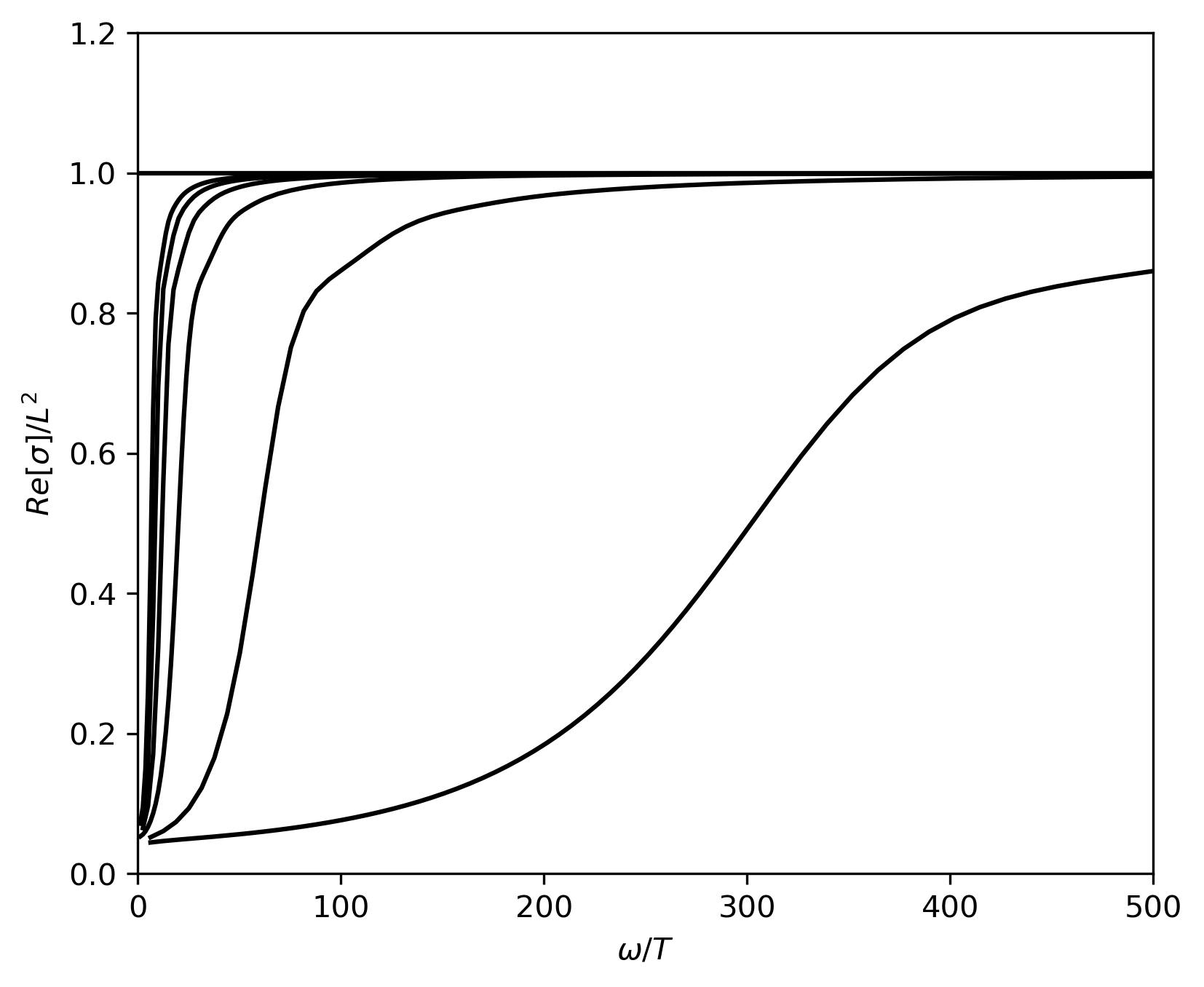}
		\label{subfig:7a}}
	\subfigure[]{\includegraphics[width=0.45\textwidth]{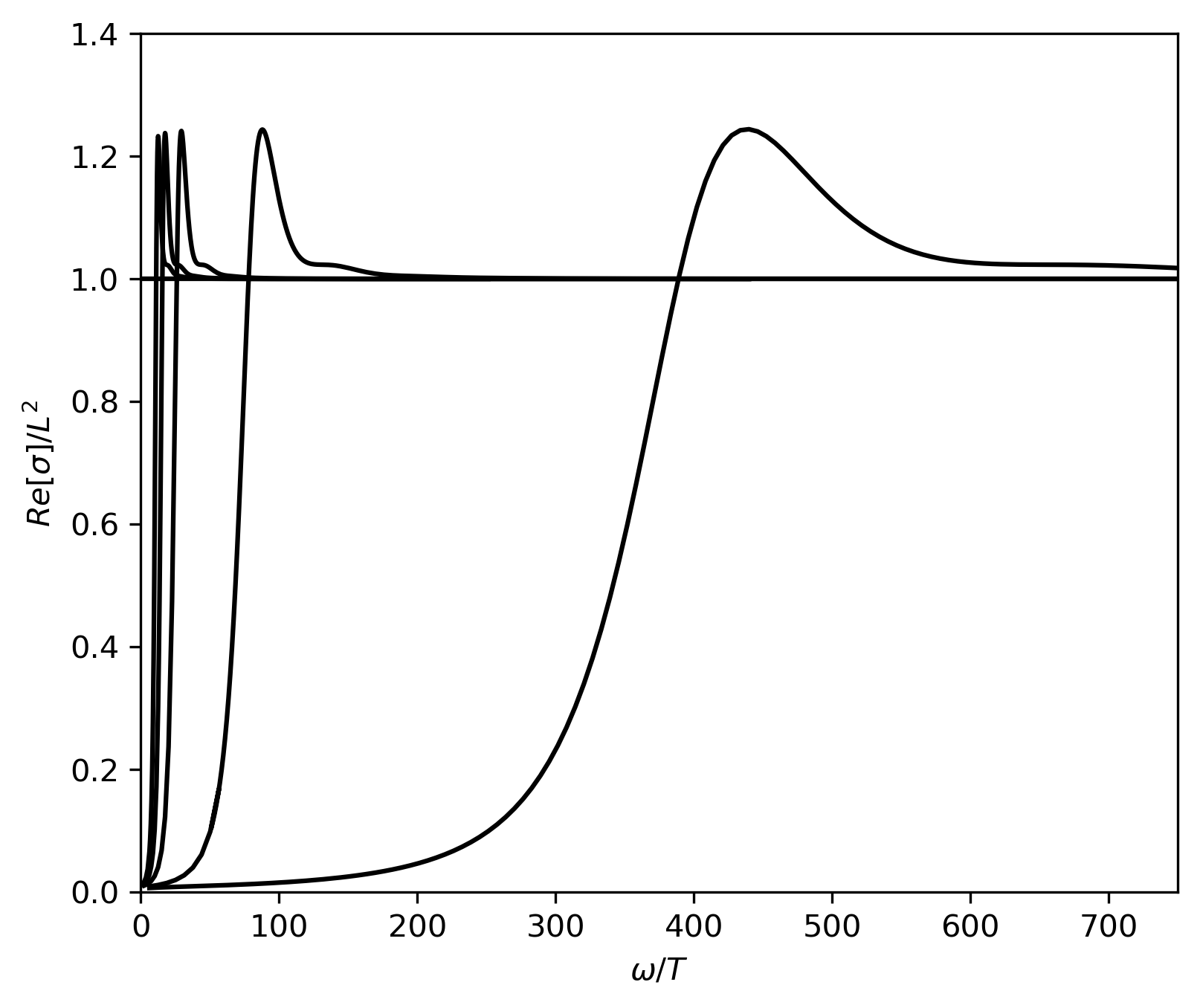}
		\label{subfig:7b}}
\caption{The evolution of Re $\sigma \left( \omega  \right)$ versus $\omega /T$ at $\varepsilon =4\pi, Q = 1 , L = 9$ for $ O_1 $ (a) and $ O_2 $ (b). 
Each graph, from left to right, is plotted for the cases $T/T_c$ = 0.7, 0.5, 0.3, 0.1 and 0.02 respectively.}
\label{fig:sigmak1L9}
\end{figure}
%

\subsubsection{$\varepsilon =16\pi,~ Q=1,L=0.5$, and $2$}

\begin{figure}
\centering
	\subfigure[]{\includegraphics[width=0.4\textwidth]{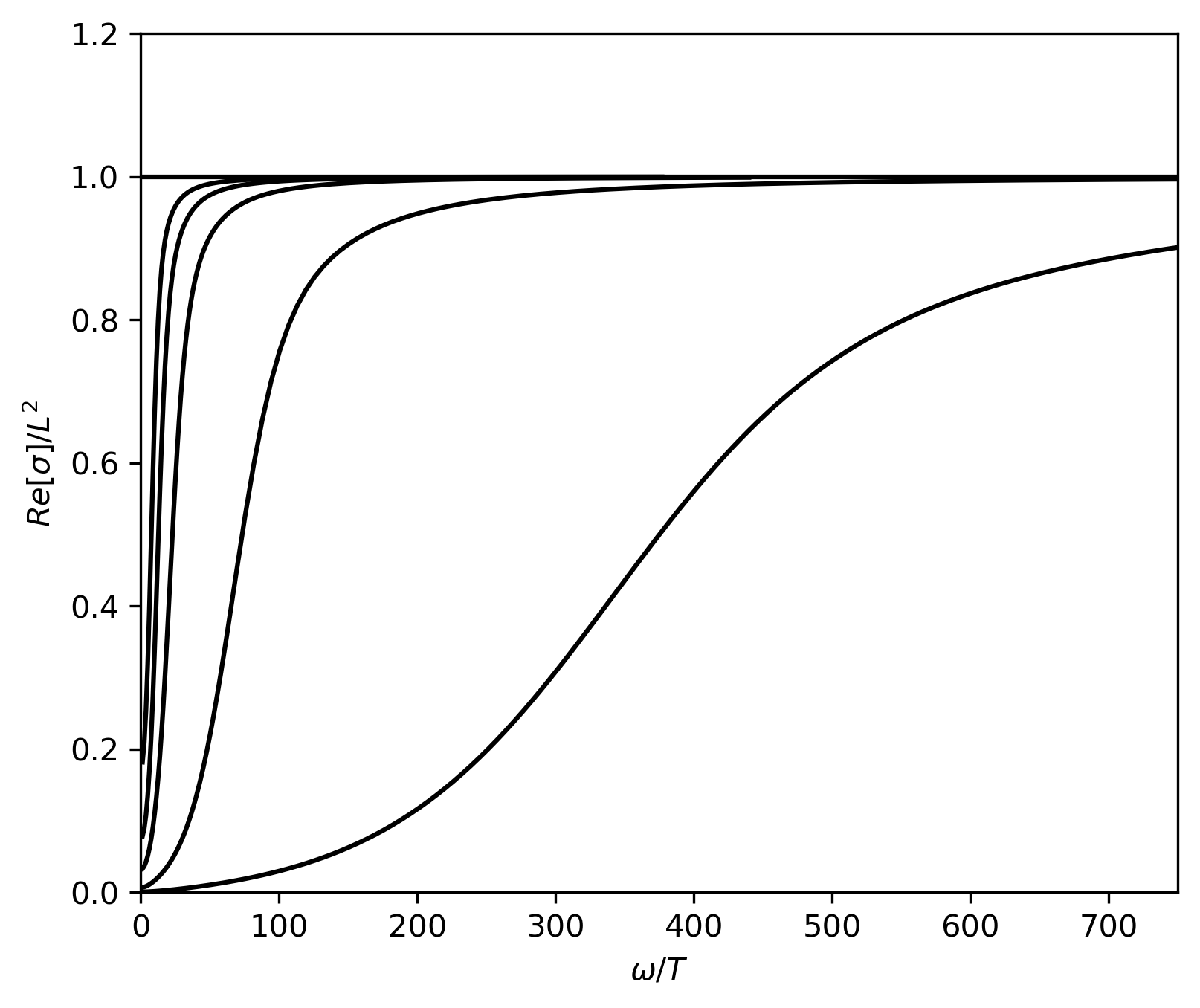}
		\label{subfig:8a}}
	\subfigure[]{\includegraphics[width=0.4\textwidth]{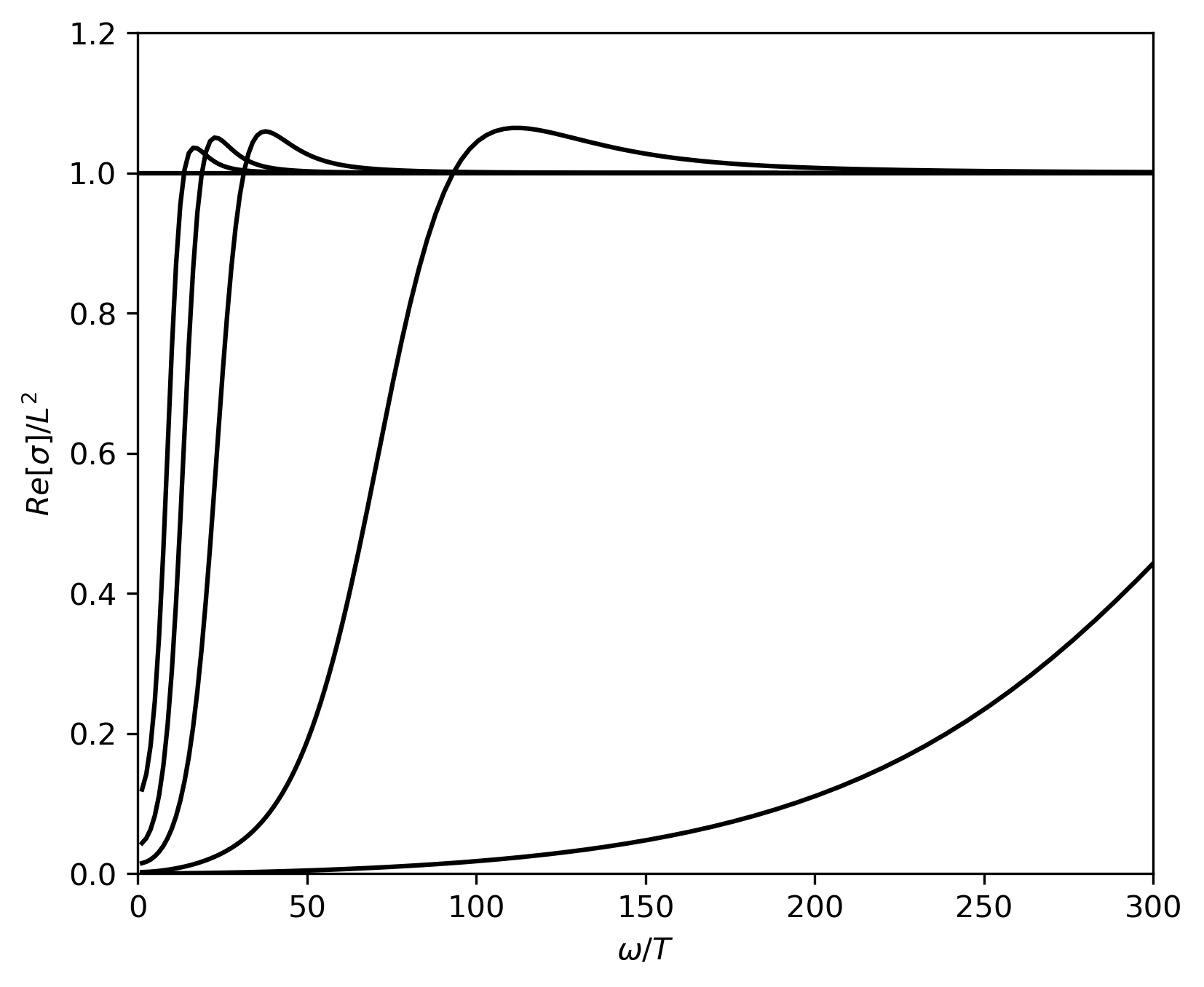}
		\label{subfig:8b}}
\caption{The evolutions of  Re $\sigma \left( \omega  \right)$versus   $\omega /T$  at $\varepsilon =16\pi, L = 0.5$ for $ O_1$ (a) and $ O_2 $ (b). 
Each graph, from left to right, is plotted for the cases $T/T_c$ = 0.7, 0.5, 0.3, 0.1 and 0.02 respectively.}
\label{fig:sigmak4L05}
\end{figure}

\begin{figure}
\centering
	\subfigure[]{\includegraphics[width=0.45\textwidth]{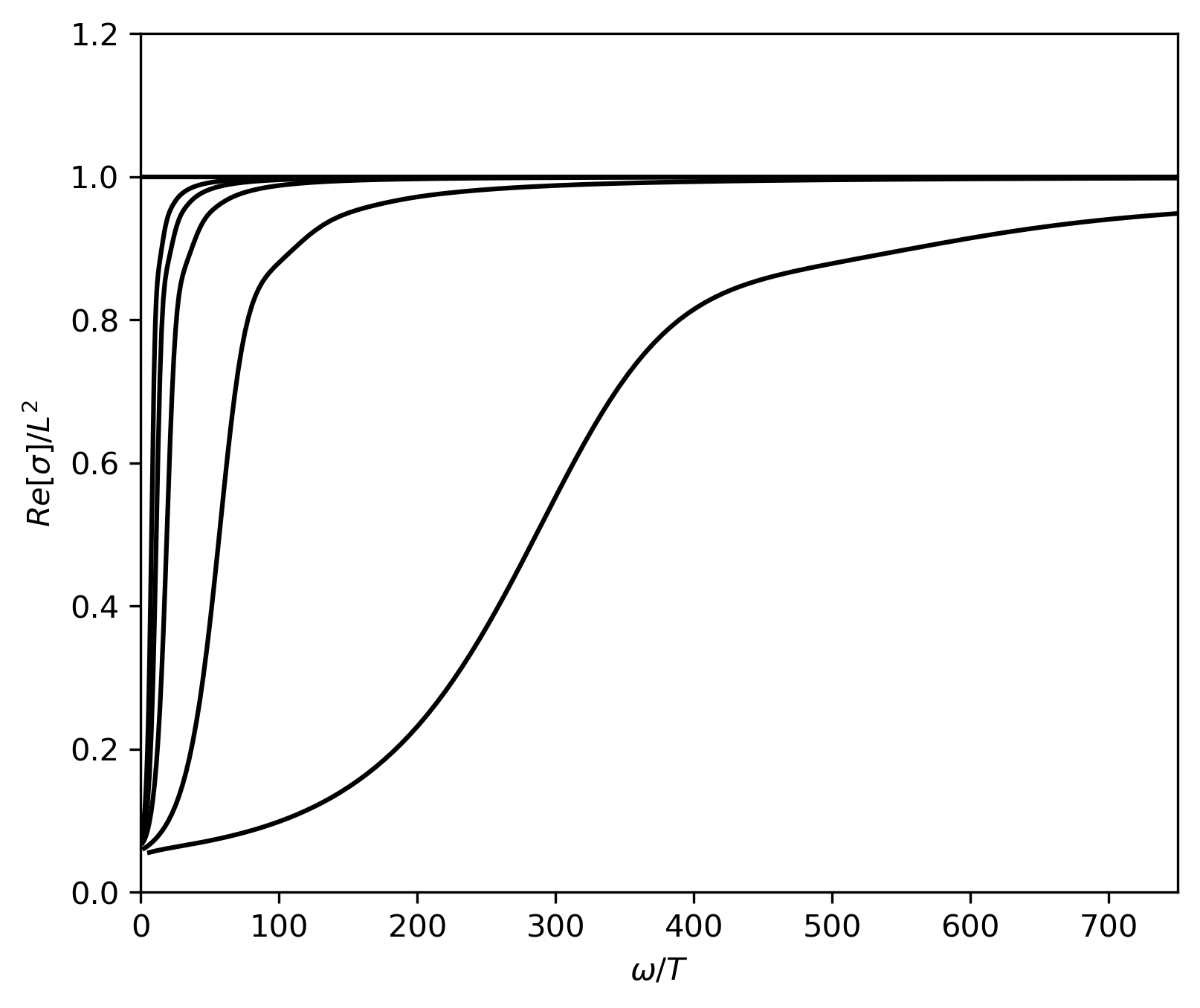}
		\label{subfig:9a}}
	\subfigure[]{\includegraphics[width=0.45\textwidth]{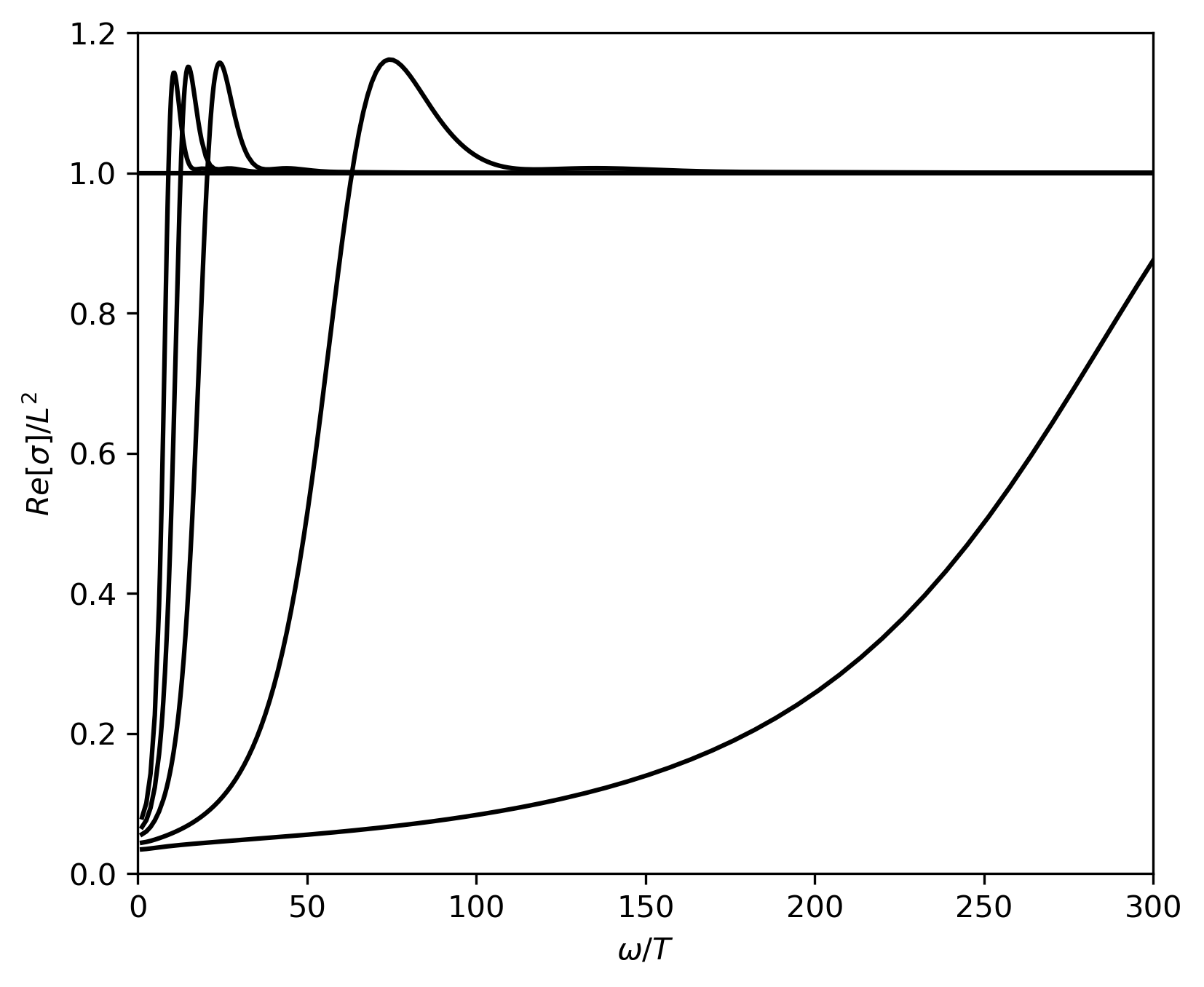}
		\label{subfig:9b}}
\caption{The evolution of Re $\sigma \left( \omega  \right)$ versus  $\omega /T$ at $\varepsilon =16\pi, L =2$  for $ O_1 $ (a) and$  O_2 $ (b). Each graph, from left to right, is plotted for the cases $T/T_c$ = 0.7, 0.5, 0.3, 0.1 and 0.02 respectively.}
\label{fig:sigmak4L20}
\end{figure}

Corresponding to $L = 0.5 $ we depict the graphs in Figs. \ref{fig:sigmak4L05}a, \ref{fig:sigmak4L05}b 
which prove clearly that the gap is really reduced even when the phase transition is second order, 
this implies that a mixture of normal and conductive states emerges. At $L = 2$ the gap totally 
disappears as seen in Figs. \ref{fig:sigmak4L20}a, \ref{fig:sigmak4L20}b, the superconductivity is totally lost. 
Thus, the conductivity  of superconductor is greatly affected  by topological charge. 
These features are similar to those of some strongly coupled superconductors \cite{c19, c20, c21}.

\section{Conclusion and outlook}\label{4}
Our paper is started from the topological black hole in which the pressure is related to cosmological constant. On this basis, we have investigated the effect on the holographic phase transitions caused by the pressure in the bulk. Our numerical computation proved that the process of condensation is favored at finite topological charge and, in particular, the pressure variation in the bulk generates a mechanism for changing the order of phase transitions in the boundary: the second order phase transitions occur at high pressures while at lower pressures they become first order. This effect has been confirmed by the free energy difference. It is clear that this mechanism manifests only in the topological AdS black hole where the pressure has been introduced through cosmological constant.

Considering the frequency dependent conductivity $\sigma \left( \omega  \right)$ we have found that  an energy gap is formed at higher pressures  and, moreover, the real and imaginary parts of $\sigma \left( \omega \right)$ fulfill the Kramers – Kronig relation. At lower pressures, however, the first order phase transitions replace the second order ones and the gap is either reduced or  disappears. This means that at small values of pressure  the superconductor is either in the mixture state of normal and superconductive states or its superconductivity is totally lost. Our set up provides a new approach to the construction of a new type of holographic superconductors in which the topological charge plays crucial role.

The analogous effect was also presented in other papers, for example, Refs. \cite{c6, c7} with  different mechanisms. In Ref. \cite{c6} the charged scalar field is forced to condense by another neutral scalar field  and in Ref. \cite{c7} the nonlinear interaction of charged scalar field was employed $F=\sum\limits_{2}^{4}{{{c}_{i}}}{{\left| \psi  \right|}^{i}}$.

One found that first order phase transitions occurred for ${{c}_{4}}\ge 1$ and, moreover, the gap becomes narrower as $ c_4 $ increases from $0$ to $ 1$.

In Ref. \cite{c22} dealing with the unbalanced Stackelberg holographic superconductors the authors showed that depending on the values of the Stackelberg model's parameters the phase transitions also change from second to first order and, at the same time, the conductivity gaps are affected strongly. 

Last but not least it is clear that the results found above allows us to have a better understanding on the physical meaning of topological charge. In this regard, we need to explore more and more the the impacts of this charge on various physical processes. The backreaction of matter fields will be the subject of our next research.

\section*{Acknowledgments}
This paper was supported by the Vietnam National Foundation for Science and Technology Development under the Grant No. 103.01-2017.300. We thank  P.H.Lien, N. T. Anh, L. V. Hoa and H. V. Quyet for useful discussions.

\end{document}